\documentclass[a4paper,11pt,titlepage]{article}

\usepackage{amssymb}
\newcommand{\eqa}[1]{\begin{eqnarray*} #1 \end{eqnarray*}}
\newcommand{\eq}[1]{\begin{equation} #1 \end{equation}}

\begin{document}

\begin{titlepage}

\rightline{NBI-HE-98-07}
\rightline{hep-th/9802190}
\rightline{February, 1998}

\vskip 1cm

\centerline{\Large \bf Coupling Constants and Brane Tensions from}
\centerline{\Large \bf Anomaly Cancellation in M-theory}

\vskip 1cm

\centerline{{\bf Troels Harmark}\footnote{e-mail: harmark@nbi.dk} }
\vskip 0.2cm
\centerline{\sl Niels Bohr Institute}
\centerline{\sl Blegdamsvej 17, DK-2100 Copenhagen \O, Denmark}

\vskip 2cm

\centerline{\bf Abstract}

The theory of eleven dimensional supergravity on 
\( \mathbb{R}^{10} \times S^1 / \mathbb{Z}_2 \) with super Yang-Mills theory on
the boundaries is reconsidered. We analyse the general solution of the
modified Bianchi identity for the four-form field strength
using the equations of motion for the three-form
and find that the four-form field strength has a unique value on the
boundaries of \( \mathbb{R}^{10} \times S^1 / \mathbb{Z}_2 \).
Considering the local supersymmetry in the ``downstairs'' approach
this leads to a relation between the eleven dimensional supergravity
coupling constants in the ``upstairs'' and ``downstairs'' approaches.
Moreover, it is shown using flux quantization that the brane tensions only 
have their standard form in the ``downstairs'' units.
We consider the gauge variation of the classical theory and find that it cannot be 
gauge invariant, contrary to a recent claim. 
Finally we consider anomaly cancellation in the ``downstairs''
and ``upstairs'' approaches and obtain the values of \( \lambda^6 / \kappa^4 \)
and the two- and five-brane tensions.

\end{titlepage}


\section{Introduction and summary}

In several recent papers\cite{horavawitten2,horavawitten,alwis2,dudasmourad,conrad,lu,faux}
the eleven dimensional supergravity theory on 
\( \mathbb{R}^{10} \times S^1 / \mathbb{Z}_2 \) with super Yang-Mills theory on
the boundaries has been discussed. Three 
important issues have come up:\newline
1) With the convention that 
\( \frac{1}{\bar{\kappa}^2} \int_{M^{11}_U} d^{11} x \frac{1}{2} \sqrt{-G} R \) 
is the Einstein-Hilbert term in the action in the ``upstairs'' approach and
\( \frac{1}{\kappa^2} \int_{M^{11}_D} d^{11} x \frac{1}{2} \sqrt{-G} R \) 
is the Einstein-Hilbert term in the action in the ``downstairs'' approach
(se next section for definition of $M^{11}_U$ and $M^{11}_D$), it was argued in 
\cite{horavawitten} that \( \bar{\kappa}^2 = 2\kappa^2 \) and that 
\( \bar{\kappa} \)
was the eleven dimensional supergravity coupling constant.
In \cite{conrad} it was argued that \( \bar{\kappa}^2 = 2\kappa^2 \), but 
that \( \kappa \) was
the eleven dimensional supergravity coupling constant. Finally, in
\cite{lu} it was argued that \( \bar{\kappa} = \kappa \) so that
both \( \kappa \) and \( \bar{\kappa} \) was the eleven dimensional
supergravity coupling constant. It is crucial which of the
coupling constants that is the right one, we have for instance
the relation\cite{alwis} 
\( 2\kappa_{11}^2 = (2\pi)^8 (\alpha')^{\frac{9}{2}} \) 
here with the name \( \kappa_{11} \) for the eleven dimensional 
supergravity coupling constant.
In this paper we prove the relation 
\( \bar{\kappa}^2 = 2\kappa^2 \) 
and that \( \kappa \) is the eleven dimensional
supergravity coupling constant.\newline
2) In \cite{faux} it was conjectured that there exists a consistent classical
theory of eleven dimensional supergravity with super Yang-Mills theory on the 
boundaries. This was based on the general solution to the Bianchi identity 
where an arbitrary parameter is introduced. For certain values of this parameter 
the theory was shown to be gauge invariant. This is completely contrary to the
original claim of Ho\u{r}ava and Witten in \cite{horavawitten} that a consistent
theory necessarily is a quantum theory, since the classical theory is not 
gauge invariant. In a quantum theory, the gauge variation of the classical theory 
can then cancel with the gauge anomaly in the effective action. 
In this paper we show that the
gauge variation of the classical theory cannot be zero, so quantization of the 
theory is necessary.\newline
3) Starting with a general solution of the modified Bianchi identity, it was
claimed in \cite{lu} that it was necessary to use gauge, gravitational and mixed
anomaly cancellation and to include two- and five-brane quantization plus a 
half-integral quantization of \( G^W / 2\pi \)\cite{witten}(see later for details) 
in order
to determine the value of \( \lambda^6 / \kappa^4 \). It was further claimed
that one could not determine \( \lambda^6 / \kappa^4 \) in the ``downstairs''
approach. In this paper we show that it is possible to determine 
\( \lambda^6 / \kappa^4 \) by working with gauge anomaly cancellation 
in the ``downstairs'' approach alone.\newline

We start in section 2 by considering the general solution to the modified Bianchi
identity where we have an arbitrary parameter called $\beta$. 
By the equations of motion for the three-form $C$ it is shown that the
four-form field strength $K$ surprisingly has a value on the boundaries
of \( \mathbb{R}^{10} \times S^1 / \mathbb{Z}_2 \) that is independent of $\beta$. 
In section 3 we consider 
the local supersymmetry of the theory in the ``downstairs'' approach and 
use this to find the value
of the four-form $K$ on the boundaries
of \( \mathbb{R}^{10} \times S^1 / \mathbb{Z}_2 \). 
Comparing the two values of $K$, we find that 
\( \bar{\kappa}^2 = 2\kappa^2 \). In section 4 we find that the
classical theory cannot be gauge invariant so that a quantized theory with
anomalies is necessary. In section 5 we consider the gauge anomaly
cancellation in the ``downstairs'' approach and find 
\( \lambda^6 / \kappa^4 = (4\pi)^5 \) and
\( \lambda^6 / \bar{\kappa}^4 = 256 \pi^5 \). In section 6 we
use this to prove that \( \kappa \) must be the
eleven dimensional supergravity coupling constant by use of
the flux quantization rule\cite{witten}. In sections 7 and 8 
we consider gauge, gravitional and mixed anomaly cancellations, in
section 7 it is in the ``downstairs'' approach and in section 8 it is in the
``upstairs'' approach. The two methods give the same results.\newline

An important conclusion to draw from this paper, is that one can derive all the
results without using the ``upstairs'' method at all. The modification of the
Bianchi identity is not necessary, since from section 3 we see that we can calculate the
four-form field strength $K$ on the boundaries by working entirely 
in the ``downstairs'' approach. This is contrary to
the ``upstairs'' approach where one has to use the four-form field strength
in the ``downstairs'' approach, 
in order to use the flux quantization rule\cite{witten}.
So the ``downstairs'' approach seems to have all the advantages: It is 
more natural conceptually, there is no need for the modification 
of the Bianchi identity
with the arbitrary parameter in the solution, and the anomaly cancellation
is easier to work out.


\section{Analysis of the modified Bianchi identity}

The ``downstairs'' eleven dimensional space-time is
\( M^{11}_D = \mathbb{R}^{10} \times S^1 / \mathbb{Z}_2 = \mathbb{R}^{10} \times 
[0,\pi\sqrt{\alpha'}] \)
and the ``upstairs'' eleven dimensional space-time is 
\( M^{11}_U = \mathbb{R}^{10} \times S^1 \) with 
\( x^{11} \mbox{ equivalent to } x^{11} + 2\pi\sqrt{\alpha'} \).
\( M^{10} \) is the boundary at $x^{11}=0$ and \( M'^{10} \) is the boundary
at $x^{11} = \pi \sqrt{\alpha'} $.\newline

The bosonic terms in the ``upstairs'' eleven dimensional supergravity theory is
\cite{cremmer,horavawitten}\footnote{We use the eleven dimensional supergravity
action with the notation from \cite{horavawitten}, except for the indices and 
the renaming of the four-form field strength.}
\[ S_{SUGRA} = \frac{1}{\bar{\kappa}^2} \int_{M^{11}_U} d^{11} x \sqrt{-G}
\Big( \frac{1}{2} R
- \frac{1}{48} K_{MN\Xi \Upsilon} K^{MN\Xi \Upsilon} \Big) + S_{CKK} \]
with
\[ S_{CKK} = - \frac{\sqrt{2}}{\bar{\kappa}^2} \int_{M^{11}_U} C\wedge K\wedge K \]
where $C$ is the three-form and $K=6dC$ is the four-form 
field strength. We use uppercase greek letters for the 11 
dimensional indices and lowercase greek letters for the
10 dimensional indices. The eleven dimensional 
metric\footnote{We use the metric signature \( -++\cdots+ \).} is $G_{MN}$ and the
ten dimensional metric is $g_{\mu \nu}$. We choose 
\( \Gamma_{\bar{11}} = \Gamma_{\bar{0}} \Gamma_{\bar{1}} \cdots \Gamma_{\bar{9}} \)
where the bars on the indices indicate flat indices.
In the ``upstairs'' approach we introduce an
orbifold transformation acting as 
\( x^{11} \rightarrow \tilde{x}^{11} = -x^{11} \) 
and \( x^\mu \rightarrow \tilde{x}^\mu = x^\mu\). If we demand that
the langrangian is invariant under this transformation we get that 
\( C_{\mu \nu \xi} (\tilde{x}) = - C_{\mu \nu \xi} (x) \),
\( C_{11 \mu \nu} (\tilde{x}) = C_{11 \mu \nu} (x) \),
\( K_{\mu \nu \xi \upsilon} (\tilde{x}) = - K_{\mu \nu \xi \upsilon} (x) \)
and \( K_{11 \mu \nu \xi} (\tilde{x}) = K_{11 \mu \nu \xi} (x) \).
We combine this action with super Yang-Mills theory
on $M^{10}$ and $M'^{10}$. On $M^{10}$ the bosonic part of the super Yang-Mills
action is
\[ S_{SYM} = -\frac{1}{\lambda^2} \int_{M^{10}} d^{10} x \sqrt{-g} \frac{1}{4} 
\mbox{tr} ( F_{\mu \nu} F^{\mu \nu} ) \]
where $F = dA + A^2$ is the gauge field strength and $A$ is the 
gauge field connection.
As Ho\u{r}ava and Witten pointed out in \cite{horavawitten}, 
local supersymmetry in the ``upstairs'' approach requires a modification
of the Bianchi-identity $dK =0$ to
\eq{ \label{bianchi1} dK = \frac{1}{\sqrt{2}} \frac{\bar{\kappa}^2}{\lambda^2}
\delta (x^{11} ) dx^{11} \wedge I_4 }
where \( I_4 = -\mbox{tr}(F^2) \). This identity has the general solution\cite{dudasmourad,lu}
\[ K = 6 dC - \frac{\beta}{\sqrt{2}} \frac{\bar{\kappa}^2}{\lambda^2} 
\delta (x^{11}) dx^{11} \wedge I_3
+ \frac{1}{2\sqrt{2}} (1-\beta) \frac{\bar{\kappa}^2}{\lambda^2} 
\epsilon(x^{11} ) I_4 \]
where $\beta$ is an arbitrary parameter and where
\( I_3 = - \mbox{tr} (AdA + \frac{2}{3}A^3) \) so that \( dI_3 = I_4 \).
In \cite{lu} Lu added the condition that $C_{\mu \nu \xi} = 0$ on $M^{10}$,
but as we shall see, this is not consistent with the equations of motion.\newline

The equations of motion for the three-form $C$ in eleven dimensional supergravity is
\eqa{ & & \partial_M ( \sqrt{-G} K^{MN\Xi \Upsilon} ) 
+ \frac{\sqrt{2}}{72} \epsilon^{M_1 M_2 \cdots M_7 MN\Xi \Upsilon}
\partial_M ( K_{M_1 M_2 M_3 M_4} C_{M_5 M_6 M_7})
\\ & & 
= \frac{\sqrt{2}}{3456} \epsilon^{M_1 M_2 \cdots M_8 N\Xi \Upsilon}
K_{M_1 M_2 M_3 M_4} K_{M_5 M_6 M_7 M_8} }
where we again only consider the bosonic terms of the action. 
We see that the only term proportional to
\( \partial_{11} K_{11 \mu \nu \xi } \) is the term 
\( \sqrt{-G} \partial_{11} K_{11 \mu \nu \xi } \), 
so if $K_{11 \mu \nu \xi}$ has a term proportional 
to $\delta(x^{11})$ then $\partial_{11} K_{11 \mu \nu \xi }$ has a term
proportional to $\partial_{11}\delta(x^{11})$. But it is not possible for
any of the other terms in the equations of motion to be proportional to
$\partial_{11}\delta(x^{11})$, so we conclude that
$K_{11 \mu \nu \xi }$ cannot have a term proportional to $\delta(x^{11})$. 
Since \( K_{11 \mu \nu \xi} (\tilde{x}) = K_{11 \mu \nu \xi} (x) \)
we cannot have a term proportional to $\epsilon (x^{11})$ 
in $K_{11 \mu \nu \xi}$ either. This means that $K_{11 \mu \nu \xi}$
is well-defined at $x^{11} = 0$, so we must have that
\[ dC_{11 \mu \nu \xi } = \frac{\beta}{6\sqrt{2}} 
\frac{\bar{\kappa}^2}{\lambda^2} \delta (x^{11}) (I_3)_{\mu \nu \xi} 
+ U_{\mu \nu \xi} \]
where $U_{\mu \nu \xi}$ is a 10 dimensional 3-form that is well-defined at 
\( x^{11} = 0 \). With the definition \( B_{\mu \nu} \equiv C_{11 \mu \nu} \)
we have
\[ dC_{11 \mu \nu \xi} = \partial_{11} C_{\mu \nu \xi}
- dB_{\mu \nu \xi} \]
but if we set 
\( dB_{\mu \nu \xi} = c \delta(x^{11}) (I_3)_{\mu \nu \xi} + U'_{\mu \nu \xi} \),
where $U'_{\mu \nu \xi}$ is a 10 dimensional 3-form without any 
terms proportional to $\delta(x^{11})$, we get 
\( d(dB)_{\mu \nu \xi \upsilon} 
= c \delta(x^{11}) (I_4)_{\mu \nu \xi \upsilon} 
+ dU'_{\mu \nu \xi \upsilon} \)
but since $d(dB)_{\mu \nu \xi \upsilon} = 0$ we must have $c=0$ 
since $dU'_{\mu \nu \xi \upsilon}$ cannot cancel a term
proportional to $\delta(x^{11})$. 
Since \( B_{\mu \nu} (\tilde{x}) = B_{\mu \nu} (x) \) there cannot be terms
in $B_{\mu \nu}$ proportional to $\epsilon(x^{11})$,
so $B_{\mu \nu}$ must be well-defined at $x^{11} = 0$.
So since
\( \partial_{11} C_{\mu \nu \xi}
= \frac{\beta}{6\sqrt{2}} 
\frac{\bar{\kappa}^2}{\lambda^2} \delta (x^{11}) (I_3)_{\mu \nu \xi} 
+ U_{\mu \nu \xi} + dB_{\mu \nu \xi} \) 
we get
\eq{ \label{c_expression} 
C = \frac{\beta}{12\sqrt{2}} \frac{\bar{\kappa}^2}{\lambda^2} 
\epsilon (x^{11}) I_3 + C' }
where $C'$ is an 11 dimensional 3-form that is well-defined at $x^{11} = 0$.
From this we obtain
\[ dC = \frac{\beta}{6\sqrt{2}} \frac{\bar{\kappa}^2}{\lambda^2} 
\delta (x^{11}) dx^{11} \wedge I_3 
+ \frac{\beta}{12\sqrt{2}} \frac{\bar{\kappa}^2}{\lambda^2}
\epsilon (x^{11}) I_4 + dC' \]
and
\eq{ \label{k_expression}
K = \frac{1}{2\sqrt{2}} \frac{\bar{\kappa}^2}{\lambda^2}
\epsilon (x^{11}) I_4 + 6dC' }
So the surprising result is that K does not have any depence on the 
arbitrary parameter $\beta$, contrary to what was found in \cite{lu,faux}.\newline

A membrane must experience the same field strength $K$ in the bulk
in the two approaches.
This means that we can find $K$ on $M^{10}$ in the ``downstairs'' approach by
taking the limiting value of $K$ for $x^{11} \rightarrow 0^+$ in 
the ``upstairs'' approach\cite{horavawitten,conrad,lu}.
This gives
\eq{ \label{k_expression2}
 K|_{M^{10}} = \frac{1}{2\sqrt{2}} \frac{\bar{\kappa}^2}{\lambda^2} I_4 }


\section{Local supersymmetry in the ``downstairs'' approach}

In \cite{horavawitten} it was shown that under a local supersymmetry 
transformation, the \( \psi \eta F^2 \) terms in the variation 
of the super Yang-Mills action on $M^{10}$ 
is\footnote{This is the variation of the modified super Yang-Mills action 
derived in \cite{horavawitten} starting from the globally supersymmetric 
super Yang-Mills action.}
\[ \Delta = \frac{1}{96\lambda^2}\int_{M^{10}} d^{10} x \sqrt{-g} 
\bar{\psi}_\mu \Gamma^{\mu \nu \xi \upsilon \rho} \eta 
\, \mbox{tr}(F^2)_{\nu \xi \upsilon \rho} \]
in the notation of \cite{horavawitten}(except for the indices). To cancel this
variation in the ``upstairs'' approach, 
one modifies the Bianchi identity for $K$ as described in the previous section.
In the ``downstairs'' approach, we must instead consider total derivatives
with respect to the eleventh coordinate in the supersymmetry
variation of the langrangian.  
We consider the following term in the eleven dimensional 
supergravity action\cite{cremmer,horavawitten}
\[ -\frac{1}{\kappa^2} \int_{M^{11}_D} d^{11} x \sqrt{-G} 
\frac{\sqrt{2}}{192} \bar{\psi}_M \Gamma^{MN\Xi\Upsilon P \Theta} 
\psi_\Theta K_{N\Xi\Upsilon P} \]
in the notation of \cite{horavawitten}(except for the indices and the renaming
of the four-form $K$). Under a supersymmetry transformation we have
that \( \delta \psi_M = \partial_M \eta + \cdots \) where \( \cdots \) represents the
terms without derivatives of \( \eta \). Ignoring the \( \cdots \)
terms we get the variation 
\[ -\frac{1}{\kappa^2} \int_{M^{11}_D} d^{11} x \sqrt{-G}
\frac{\sqrt{2}}{96} \bar{\psi}_M \Gamma^{MN\Xi\Upsilon P \Theta} 
\partial_{\Theta} \eta K_{N\Xi\Upsilon P} \]
so the contribution from this containing a total eleventh derivative is
\[  -\frac{1}{\kappa^2} \int_{M^{11}_D} d^{11} x \partial_{11} ( \sqrt{-G}
\frac{\sqrt{2}}{96} \bar{\psi}_\mu \Gamma^{\mu \nu \xi \upsilon \rho 11} 
\eta K_{\nu \xi \upsilon \rho} ) \]
Ignoring the contribution from $M'^{10}$ this becomes
\[ \frac{1}{\kappa^2} \int_{M^{10}} d^{10} x \sqrt{-g}
\frac{\sqrt{2}}{96} \bar{\psi}_\mu \Gamma^{\mu \nu \xi \upsilon \rho}
\eta K_{\nu \xi \upsilon \rho} \]
where we used that \( \Gamma^{11} \eta = \frac{\sqrt{-g}}{\sqrt{-G}} \eta \) on \(M^{10}\).
In order to cancel the $\Delta$ term, we must require
\eq{ \label{k_expression3}
 K|_{M^{10}} = - \frac{1}{\sqrt{2}} \frac{\kappa^2}{\lambda^2} 
\mbox{tr}(F^2) = \frac{1}{\sqrt{2}} \frac{\kappa^2}{\lambda^2} I_4 }
This is is completely consistent with (\ref{k_expression2}) from the ``upstairs''
approach provided that we have the relation
\[ \bar{\kappa}^2 = 2\kappa^2 \]
So this relation can be seen as a consequence of demanding local supersymmetry
in both the ``upstairs'' and the ``downstairs'' approach.


\section{The gauge variation of the classical theory}

In the following we calculate the gauge variation of the classical theory.
In the ``downstairs'' approach we always have $K=6dC$ so we have that
\[ C|_{M^{10}} = \frac{1}{6\sqrt{2}} \frac{\kappa^2}{\lambda^2} I_3 \]
up to an irrelevant exact form. If we now make the gauge variation
\( \delta A = [D,v] = dv + [A,v] \) we have that $\delta K|_{M^{10}} = 0$ and
\[ \delta C|_{M^{10}} = \frac{1}{6\sqrt{2}} \frac{\kappa^2}{\lambda^2} dI^1_2 \]
where $I^1_2 = - \mbox{tr} (vdA)$. The only possible non-gauge-invariant term in the
combined supergravity and super Yang-Mills action is the $S_{CKK}$ term, and we have
\eqa{& &
\delta S_{CKK} = -\frac{\sqrt{2}}{\kappa^2} \int_{M^{11}_D} 
\delta C \wedge K \wedge K 
= - \frac{1}{6\lambda^2} \int_{M^{11}_D} d( I^1_2 \wedge K \wedge K )
\\ & &
= \frac{1}{6\lambda^2} \int_{M^{10}} I^1_2 \wedge K \wedge K
= \frac{1}{12} \frac{\kappa^4}{\lambda^6} \int_{M^{10}} I^1_2 \wedge (I_4)^2 }
So the classical theory cannot be gauge-invariant, contrary to the claim in \cite{faux}. 
As explained in \cite{horavawitten}, 
we must then consider the combined eleven dimensional supergravity and super
Yang-Mills theory as an effective low-energy theory for a quantum theory, so that
the quantum anomalies in the effective action can cancel the gauge variation of the
classical theory.


\section{The gauge anomaly in the ``downstairs'' approach}

In order to cancel the gauge variation of the classical theory, we must use the gauge group
$E_8$\cite{horavawitten}. For ten dimensional Majorana-Weyl spinors, we have the 
12-form(see \cite{conrad}; the spinors have positive chirality 
under \( \Gamma_{\bar{11}} \))
\[ I_{12} = - \frac{1}{384(2\pi)^5} (I_4)^3 \]
with the descent equations
\[ I_{12} = dI_{11},\ \ \delta I_{11} = dI^1_{10} \]
so that the gauge anomaly in the effective action is
\[ \delta W = \int_{M^{10}} I^1_{10} 
= - \frac{1}{384(2\pi)^5} \int_{M^{10}} I^1_2 \wedge (I_4)^2 \]
Since we want \( \delta S_{CKK} + \delta W = 0 \) we find
\[ \frac{\lambda^6}{\kappa^4} = (4\pi)^5,\ \ 
\frac{\lambda^6}{\bar{\kappa}^4} = 256 \pi^5 \]
This is the same result as in \cite{conrad}. 
It is important to note that we obtained this by using only the gauge 
anomaly in the ``downstairs'' approach. This was deemed impossible in \cite{lu}.\newline


\section{Proof that $\kappa$ is the eleven dimensional supergravity 
coupling constant}

We have proved that \( \bar{\kappa}^2 = 2\kappa^2 \). To prove that 
\( \kappa \) is the eleven dimensional supergravity coupling constant we
start by assuming that \( \bar{\kappa} \) is the eleven 
dimensional supergravity coupling constant and show that this leads to
an inconsistency. From \cite{alwis} we have the quantization rule for
the two-brane tension
\[ (T_2)^3 = \frac{(2\pi)^2}{2\bar{\kappa}^2 m},\ m \in \mathbb{Z} \]
From \cite{witten} we know that \( G^W /2\pi \) should have a half
integral period, where in our notation \( G^W = \sqrt{2}T_2 K \), 
so since\footnote{In sections 7 and 8 we extend our anomaly analysis 
to include gravitational 
and mixed anomalies. This has the consequence that \( I_4 = - \mbox{tr}(F^2) \)
is replaced by \( \hat{I}_4 = \frac{1}{2} \mbox{tr}(R^2) - \mbox{tr}(F^2) \)
in (\ref{k_expression2}) and (\ref{k_expression3}). In the following we use 
these modified expressions for the four-form field strength on \( M^{10} \).}
\[ \frac{\sqrt{2}}{2\pi} T_2 K|_{M^{10}} 
= T_2 \frac{\bar{\kappa}^2}{\lambda^2} \frac{1}{4\pi} \hat{I}_4 
= \frac{1}{(2m)^{\frac{1}{3}}} \frac{1}{16\pi^2} \hat{I}_4 \]
this means that we have the flux quantization 
rule\footnote{See also \cite{alwis2,conrad,lu}.} 
\[ \frac{1}{(2m)^\frac{1}{3}} \in \mathbb{Z} \]
but that is impossible. 
If instead we assume that \( \kappa \) is the eleven 
dimensional supergravity coupling constant, we can again write
the two-brane tension quantization rule\cite{alwis}
\[ (T_2)^3 = \frac{(2\pi)^2}{2\kappa^2 m},\ m \in \mathbb{Z} \]
giving
\[ \frac{\sqrt{2}}{2\pi} T_2 K|_{M^{10}}
= T_2 \frac{\kappa^2}{\lambda^2} \frac{1}{2\pi} \hat{I}_4
= \frac{1}{m^{\frac{1}{3}}} \frac{1}{16\pi^2} \hat{I}_4 \]
so that the flux quantization rule \cite{witten} implies
\[ \frac{1}{m^{\frac{1}{3}}} \in \mathbb{Z} \]
This is fulfilled if and only if $m=1$\footnote{Assuming $m$ is positive.} so that
\[ (T_2)^3 = \frac{(2\pi)^2}{2\kappa^2} \]
This is the standard form for the two-brane tension(see for example \cite{alwis}).


\section{Gauge, gravitational and mixed anomalies in the ``downstairs'' approach}

To extend our anomaly analysis to include gravitational and mixed anomalies, we must
replace \( I_4 \) with 
\( \hat{I}_4 = \frac{1}{2} \mbox{tr}(R^2) - \mbox{tr}(F^2) \)
in (\ref{bianchi1}) and (\ref{k_expression3}). This was pointed
out in \cite{horavawitten} based on the knowledge of the structure of ten dimensional
anomalies. Here \( R = d\omega + \omega^2 \) is the curvature two-form and \( \omega \)
is the spin connection. 
With the local Lorentz variation \( \delta \omega = [D,\Theta] \), we can write
\[ \hat{I}_3 = \frac{1}{2} \mbox{tr}(\omega d\omega + \frac{2}{3} \omega^3)
- \mbox{tr}(AdA+\frac{2}{3} A^3),\ \ \hat{I}^1_2 = \frac{1}{2}\mbox{tr}(\Theta d\omega)
- \mbox{tr}(vdA) \]
so that we have the descent equations
\[ d\hat{I}_3 = \hat{I}_4,\ \ \delta \hat{I}_3 = d\hat{I}^1_2 \]
Replacing $I$ with $\hat{I}$ we have
\[ \delta K|_{M^{10}} = 0,\ \ \delta C|_{M^{10}} = \frac{1}{6\sqrt{2}} 
\frac{\kappa^2}{\lambda^2} d\hat{I}^1_2 \]
and
\[ \delta S_{CKK} = \frac{1}{12} \frac{\kappa^4}{\lambda^6} 
\int_{M^{10}} \hat{I}^1_2 \wedge (\hat{I}_4)^2 \]
The anomalous 12-form with gauge, gravitional and mixed anomalies 
is(see \cite{conrad})
\[ \hat{I}_{12} = - \frac{1}{96(2\pi)^5} \hat{I}_4 \wedge 
( \frac{1}{4} (\hat{I}_4)^2 - X_8 ) \]
where \( X_8 = -\frac{1}{8}\mbox{tr}(R^4) + \frac{1}{32} (\mbox{tr}(R^2))^2 \),
with the descent equations
\[ \hat{I}_{12} = d\hat{I}_{11},\ \ \delta \hat{I}_{11} = d\hat{I}^1_{10} \]
so that the anomaly takes the form
\[ \delta W = \int_{M^{10}} \hat{I}^1_{10} 
= - \frac{1}{96(2\pi)^5} \int_{M^{10}} \hat{I}^1_2 \wedge (\frac{1}{4} (\hat{I}_4)^2 - X_8) \]
This only partly cancels with $\delta S_{CKK}$, so we have to introduce
the five-brane term in the classical action. In ``downstairs'' units, 
it is\cite{conrad}
\[ S_{5} = \frac{\sqrt{2}}{8(2\pi)^3 \kappa^2 T_5} \int_{M^{11}_D} C \wedge X_8 \]
This term has the variation
\[ \delta S_{5} 
= \frac{\sqrt{2}}{8(2\pi)^3 \kappa^2 T_5} \frac{1}{6\sqrt{2}} 
\frac{\kappa^2}{\lambda^2} \int_{M^{11}_D} d( \hat{I}^1_2 \wedge X_8 )
= - \frac{1}{48(2\pi)^3 \lambda^2 T_5} \int_{M^{10}} \hat{I}^1_2 \wedge X_8 \]
So we obtain \( \delta W + \delta S_{CKK} + \delta S_5 = 0 \) if and only if
\[ \frac{\lambda^6}{\kappa^4} = (4\pi)^5 \mbox{ and } 
(T_5)^3 = \frac{2\pi}{(2\kappa^2)^2} \]
We see that $T_5$ has the standard form(see for example \cite{alwis}).


\section{Gauge, gravitational and mixed anomalies in the ``upstairs'' approach}

In this section the anomaly cancellation in the ``upstairs'' approach 
is considered. The purpose is to check whether the
anomalies cancel for the same values of $\lambda^6 / \kappa^4$ and $T_5$ 
as in the ``downstairs'' method.\newline

From (\ref{c_expression}) and (\ref{k_expression}) we have
\[ C = \frac{\beta}{12\sqrt{2}} \frac{\bar{\kappa}^2}{\lambda^2} 
\epsilon (x^{11}) \hat{I}_3 + C' \]
\[ K = \frac{1}{2\sqrt{2}} \frac{\bar{\kappa}^2}{\lambda^2}
\epsilon (x^{11}) \hat{I}_4 + 6dC' \]
The variation of $C$ is 
\[ \delta C = \frac{\beta}{12\sqrt{2}} \frac{\bar{\kappa}^2}{\lambda^2} 
\epsilon (x^{11}) d\hat{I}^1_2 + \delta C' \]
Since we want $\delta K = 0$, we require \( d ( \delta C' ) = 0 \). Using
\[ K \wedge K = \frac{1}{8} \frac{\bar{\kappa}^4}{\lambda^4} (\hat{I}_4)^2 
\epsilon(x^{11})^2 + 36 dC' \wedge dC' + \frac{6}{\sqrt{2}} \frac{\bar{\kappa}^2}{\lambda^2}
\epsilon(x^{11}) \hat{I}_4 \wedge dC' \]
we find
\[ \int_{M^{11}_U} \delta C' \wedge K \wedge K 
= \frac{6}{\sqrt{2}} \frac{\bar{\kappa}^2}{\lambda^2} \int_{M^{11}_U} \epsilon(x^{11}) \hat{I}_4 
\wedge dC' \wedge \delta C' = 0 \]
where we used that $C'_{\mu \nu \xi} = 0$ on $M^{10}$ since
$C'_{\mu \nu \xi}$ is odd under the
orbifold transformation. So
\eqa{& &
\delta S_{CKK} = -\frac{\sqrt{2}}{\bar{\kappa}^2} \int_{M^{11}_U} (\delta C-\delta C')
\wedge K \wedge K
= - \frac{\beta}{12} \frac{1}{\lambda^2} \int_{M^{11}_U} \epsilon(x^{11}) d\hat{I}^1_2 
\wedge K \wedge K
\\ & &
= \frac{\beta}{6} \frac{1}{\lambda^2} \int_{M^{11}_U} \Big( \delta(x^{11}) dx^{11} \wedge
\hat{I}^1_2 \wedge K \wedge K + \epsilon(x^{11}) \hat{I}^1_2 \wedge dK \wedge K \Big)
\\ & &
= \frac{\beta}{16} \frac{\bar{\kappa}^4}{\lambda^6} \int_{M^{11}_U} \epsilon(x^{11})^2 
\delta(x^{11}) dx^{11} \wedge \hat{I}^1_2 \wedge (\hat{I}_4)^2 
= \frac{\beta}{48} \frac{\bar{\kappa}^4}{\lambda^6} \int_{M^{11}_U}  
\delta(x^{11}) dx^{11} \wedge \hat{I}^1_2 \wedge (\hat{I}_4)^2
\\ & & 
= \frac{\beta}{48} \frac{\bar{\kappa}^4}{\lambda^6} \int_{M^{10}} \hat{I}^1_2 
\wedge (\hat{I}_4)^2 } 
The five-brane term in ``upstairs'' units is
\[ S_5 = \frac{\sqrt{2}}{8(2\pi)^3 \bar{\kappa}^2 T_5} \int_{M^{11}_U} C \wedge X_8 \]
Hence
\eqa{& &
\delta S_5 = \frac{\sqrt{2}}{8(2\pi)^3 \bar{\kappa}^2 T_5} 
\frac{\beta}{12\sqrt{2}} \frac{\bar{\kappa}^2}{\lambda^2} 
\int_{M^{11}_U} \epsilon(x^{11}) d\hat{I}^1_2 \wedge X_8 
= \frac{\beta}{96(2\pi)^3 \lambda^2 T_5} \int_{M^{11}_U} 
\epsilon(x^{11}) d\hat{I}^1_2 \wedge X_8 
\\ & &
= - \frac{\beta}{48(2\pi)^3 \lambda^2 T_5} \int_{M^{11}_U} \delta(x^{11}) dx^{11} 
\wedge \hat{I}^1_2 \wedge X_8 
= - \frac{\beta}{48(2\pi)^3 \lambda^2 T_5} \int_{M^{10}} \hat{I}^1_2 \wedge X_8 }
So the anomaly cancellation \( \delta W + \delta S_{CKK} + \delta S_5 = 0 \) implies
\[ \frac{\lambda^6}{\kappa^4} = \beta (4\pi)^5 \mbox{ and }
(T_5)^3 = \beta^2 \frac{2\pi}{(2\kappa^2)^2} \]
We have the quantization rules for the two- and five-brane tensions\cite{duff,alwis}
\[ 2\kappa^2 T_2 T_5 = 2\pi n,\ n \in \mathbb{Z} \mbox{  and  }
(T_2)^3 = \frac{(2\pi)^2}{2\kappa^2 m},\ m \in \mathbb{Z} \]
So using this, we find
\[ \beta^2 = m n^3 \]
Since
\[ \frac{\sqrt{2}}{2\pi} T_2 K|_{M^{10}} 
= \frac{1}{(\beta m)^{\frac{1}{3}}} \frac{1}{16\pi^2}\hat{I}_4
= \frac{1}{\sqrt{mn}} \frac{1}{16\pi^2} \hat{I}_4 \]
we have the flux quantization rule\cite{witten}
\[ \frac{1}{\sqrt{mn}} \in \mathbb{Z} \]
so that the unique solution is\footnote{Assuming $m$ and $n$ are positive.}
\[ \beta = m = n = 1 \]
which gives
\[ \frac{\lambda^6}{\kappa^4} = (4\pi)^5,\ \ 
(T_2)^3 = \frac{(2\pi)^2}{2\kappa^2} \mbox{ and } 
(T_5)^3 = \frac{2\pi}{(2\kappa^2)^2} \]
So we get exactly the same solution as for the ``downstairs'' method.


\section*{Acknowledgements}

I would like to thank Paolo Di Vecchia and Jens Lyng Petersen for many
discussions and for reading the manuscript.


\end{document}